\documentstyle[referee]{mn}

\def\himsun{{h^{-1}M_\odot}}
\def\msun{{M_\odot}}

\def\mpc{\,h^{-1}{\rm {Mpc}}}
\def\kpc{\,h^{-1}{\rm {kpc}}}

\input epsf
\title[]
{The Growth and Structure of Dark Matter Haloes}
\author[Zhao et al.]
{D.H. Zhao$^{1,2}$, H.J. Mo$^2$, Y.P. Jing$^{1,2}$,
G. B\"orner$^{2}$ \\
$^1$ Shanghai Astronomical Observatory, the
Partner Group of MPI f\"ur Astrophysik, 
Nandan Road 80, Shanghai
200030, China\\
$^2$ Max-Planck Institut f\"ur Astrophysik,
Karl-Schwarzschild-Strasse 1, 85748 Garching, Germany\\
dhzhao@center.shao.ac.cn, hom@mpa-garching.mpg.de\\
ypjing@center.shao.ac.cn, grb@mpa-garching.mpg.de}
\date{Accepted ........
      Received .......;
      in original form .......}
\pubyear{2000}

\begin{document}
\maketitle
\begin{abstract}

In this paper, we analyse in detail the mass-accretion histories and
structural properties of dark haloes in high-resolution N-body
simulations. We model the density distribution in individual haloes
using the NFW profile.  For a given halo, there is a tight correlation
between its inner scale radius $r_s$ and the mass within it, $M_s$,
for all its main progenitors. Using this correlation, one can predict
quite well the structural properties of a dark halo at any time in its
history from its mass accretion history, implying that the structure
properties and the mass accretion history are closely correlated. The
predicted growing rate of concentration $c$ with time tends to
increase with decreasing mass accretion rate.  The build-up of dark
haloes in CDM models generally consists of an early phase of fast
accretion [where the halo mass $M_h$ increases with time much faster
than the expansion rate of the universe] and a late phase of slow
accretion [where $M_h$ increases with time approximately as the
expansion rate]. These two phases are separated at a time when
$c\sim 4$ and the typical binding energy of the halo is approximately
equal to that of a singular isothermal sphere with the same circular
velocity. Haloes in the two accretion phases show systematically
different properties, for example, the circular velocity $v_h$
increases rapidly with time in the fast accretion phase but remain
almost constant in the slow accretion phase, the inner properties of a
halo, such as $r_s$ and $M_s$ increase rapidly with time in the fast
accretion phase but change only slowly in the slow accretion phase,
the inner circular velocity $v_s$ is approximately equal to $v_h$ in
the fast accretion phase but is larger in the slow accretion phase.
The potential well associated with a halo is built up mainly in the
fast accretion phase, while a large amount of mass can be accreted in
the slow accretion phase without changing significantly the potential
well. We discuss our results in connection to the formation of dark
haloes and galaxies in hierarchical models.
\end{abstract}



\section{Introduction}

 The Cold Dark Matter (CDM) cosmogony 
(Peebles 1982; Blumenthal et al. 1984; Davis et al. 1985) 
has now been proved to be a remarkably successful 
framework for understanding the structure formation 
in the universe. In the CDM scenario, a key concept in 
the build-up of structure is the formation of dark matter haloes,
which are quasi-equilibrium systems of dark matter particles
formed through non-linear gravitational collapse.
In a hierarchical scenario like the CDM, most mass in the
Universe at any time is in dark haloes; galaxies and 
other luminous objects are assumed to form by cooling and 
condensation of baryons within these haloes (White \& Rees 1978). 
In this case, a detailed understanding of the formation and 
structure of dark matter haloes is of fundamental importance 
for predicting the properties of luminous objects.

One of the most important properties of the halo
population is their density profiles. Based on N-body simulations, 
Navarro, Frenk, \& White (1997; hereafter NFW), 
found that CDM haloes can be fitted by a 
two-parameter profile:
\begin{equation}
\label{eq:nfw}
\rho(r) = \frac{4\rho_s}{(r/r_s)\left(1+r/r_s\right)^2},
\end{equation}
where $r_s$ is a characteristic ``inner'' radius at which   
the logarithmic density slope is $-2$, and $\rho_s$ 
is the density at $r_s$. A halo is often defined so that
the mean density $\rho_h$ within the halo radius $r_h$ 
is a constant ($\Delta_h$, will be defined in Section 2)
times the mean density 
of the universe (${\bar \rho}$) at the redshift ($z$) in 
consideration. The halo mass can then be written as
\begin{equation}
\label{eq:mh}
 M_h \equiv \frac{4 \pi}{3} \Delta_h {\bar \rho} r_h^3.
\end{equation}
We define the circular velocity of a halo as
$v_h=(GM_h/r_h)^{1/2}$, and so
\begin{equation}
M_h={v_h^2 r_h\over G}
   ={2^{1/2} v_h^3\over [\Delta_h\Omega(z)]^{1/2} H(z)}\,,
\end{equation}
where $H(z)$ is the Hubble constant, and $\Omega(z)$
the mass density parameter, at redshift $z$. 
NFW introduced an alternative parameter, the concentration 
parameter $c$, defined as $c \equiv r_h/r_s$, to describe 
the shape of the halo profile. It is then easy to show that 
\begin{equation}
\label{eq:rhosc}
  \rho_s=\rho_h \frac {c^3}{12\left[\ln (1+c)-c/(1+c)\right]}.
\end{equation}
We denote the mass within $r_s$ by $M_s$, and the circular 
velocity at $r_s$ by $v_s$. These quantities are related
to $c$ and $M_h$ as
\begin{equation}
M_s={\ln 2-1/2\over \ln(1+c)-c/(1+c)}M_h\,,
~~~~~
v_s^2=v_h^2 {cM_s\over M_h}\,.
\end{equation}

In general, the structure of a halo is expected to 
depend not only on cosmology and power spectrum, but
also on its formation history. There are therefore 
attempts to relate halo concentration to other quantities
characterizing the formation of a halo.    
In their original paper, NFW suggested that the
characteristic density of a halo, $\rho_s$, should be 
a constant ($k$) times the mean density of the universe, 
${\bar \rho (z_f)}$, at the redshift $z_f$ 
(referred as the formation time of the halo by NFW)
when half of the halo's mass was first in progenitors more massive 
than $f$ times the halo mass.  NFW used the extended 
Press-Schechter formula to calculate
$z_f$ and found that the anti-correlation between
$c$ and $M_h$ observed in their simulations can be reproduced 
with a proper choice of the values for the constants $k$ and $f$. 

Subsequent analyses showed that additional complexities may be
involved in halo structure. First, it has been shown that haloes of
fixed mass have significant scatter in their $c$ values (Jing 2000),
although there is a mean trend of $c$ with $M_h$. If this trend is
indeed due to a correlation between concentration and formation time,
the scatter in $c$ may reflect the expected scatter in the formation
time for haloes of a given mass. Second, Bullock et al. (2001,
hereafter B01) found that the halo concentration (at fixed mass) is
systematically lower at higher redshift, with a trend $c \propto a$
much stronger than that predicted by the NFW model. They presented an
empirical model that can reproduce the concentration in
the LCDM model better than the original model proposed by NFW. Using the
same simulations as B01, Wechsler et al. (2002; hereafter W02) found
that, over a large mass range the mass accretion histories of
individual haloes are reasonably well described by a one-parameter
exponential form, $M_h(z)=M_h(0){\rm exp}[-2z/(1+z_f)]$. They defined
the formation time of a halo to be $z_f$. Assuming that $c$ equals to
$4.1$ at the formation time and grows proportionally to the scale
factor $a$, W02 proposed a recipe to predict the concentration $c$ for
individual haloes through their mass-accretion histories. This recipe
can reproduce the dependences of $c$ on both mass and redshift found in
B01.

Semi-analytical modeling of galaxy formation becomes a powerful tool
to predict observational properties of galaxies in hierarchical
clustering models (e.g. Kauffmann et al. 1993; Cole et al. 1994; Mo,
Mao, \& White 1998; Somerville \& Primack 1999; Mo \& Mao 2002). In the
theory, one needs to know both the mass accretion history and the
structure properties of the host halo, as they critically influence
many observable properties of the galaxy formed within it. The mass
accretion history is understood relatively well, and can be modeled
accurately either in analytical methods (e.g. extended Press-Schechter
theory, EPS) or in N-body simulations. The internal structure of a
halo at higher redshift (i.e. its main progenitor) is modeled much
more poorly, since the EPS theory tells nothing about halo internal
structure, and a high resolution is required to resolve the halo
internal structure in N-body simulations (especially for conventional
cosmological simulations). It is therefore very important to find a
way (analytical or empirical) to predict the structural history of a
halo.

The result of W02 is encouraging, because it implies that halo
structure is closely correlated with mass-accretion history. This
correlation is important not only for predicting the halo
concentration $c$, but also for understanding the formation of dark
matter haloes in hierarchical models. However, as noticed by W02
themselves, not all mass-accretion histories are smooth and have the
same form; the fit to the analytical form may be sensitive to the
presence of major mergers. Moreover, the assumption that $c \propto a$
is not expected to be universally true, because in general $c$ should
depend on the mass-accretion history even after the formation
time. Furthermore, Jing \& Suto (2002) found that for the standard CDM
model with $\Omega_m=1$, $c$ grows much faster for a fixed mass, $c
\propto a^{3/2}$, and indeed a scaling, $c \propto H^{-1}(z)$ [where
$H(z)$ is the Hubble constant at redshift $z$] may work better.
For the flat universe considered here,
$H(z)=H_0[\Omega_0(1+z)^3+\lambda_0]^{1/2}$, where $H_0$ is the
current Hubble constant.

In this paper, we analyse in detail the mass-accretion histories and
structural properties of a few high-resolution dark haloes of galactic
masses. We find that, for a given halo, there is a tight correlation
between the scale radius $r_s$ and the mass $M_s$ within it for all
its main progenitors (see Section 2 for definition).  We show that
this relation can be used to predict the structural properties (such
as $c$) of haloes at any given time from their mass-accretion
histories.  We also find several interesting properties of the mass
accretion process, which can help us to understand the formation of
dark haloes in hierarchical models. Compared with the work of W02, we
will focus on our new finding of the scaling relation between $r_s$
and $M_s$, and emphasize how to use this relation to predict the
structure properties of halos from their merger history. Our halo
sample is small but has a higher mass resolution, compared with the
halo sample of W02. We will use the GIF simulations (Kauffmann et
al. 1999), which are similar to the simulations of W02, to verify our
conclusions drawn from the small set of high-resolution
simulations. The arrangement of the paper is as follows. We describe
the simulations used in this paper and the analyses performed on them
in Section 2. The mass-accretion histories and their relations to halo
internal properties are analysed in Section 3. In Section 4 we propose
a recipe to predict halo structural properties through halo
mass-accretion histories and test it against various simulation
results. Finally, in Section 5 we discuss the implication of our
results for galaxy formation and summarize our main conclusions.  We
use physical lengths instead of comoving lengths throughout the paper
unless otherwise specified.

\section {Simulations and analyses}
In this paper, we first analyze a set of five high-resolution halo
simulations. The haloes were selected from a cosmological P$^3$M
N-body simulation for a $\Lambda$CDM model with $\Omega_0=0.3$,
$\lambda_0=0.7$, $h=0.7$ and $\sigma_8=1.0$. This simulation uses
$128^3$ particles in a comoving $(50h^{-1}{\rm Mpc})^3$ cube, and the
particle mass is $7\times10^{9}\himsun$. The selected haloes consist
of 1000 particles, have a mass of $7\times10^{12}\himsun$, and are
relatively isolated [i.e. there is no companion of larger mass within
a distance of $3(r_{\rm vir1}+r_{\rm vir2})$, where $r_{\rm vir1}$ and
$r_{\rm vir2}$ are the virial radii of the halo and the companion,
respectively, and the definition for the virial radius will be given
shortly].  These haloes are then resimulated with the multiple-grid
${\rm P^3M}$ code of Jing and Suto (2000). We use $\sim 5.5\times
10^5$ particles, each of mass $m_p \sim 1.6\times 10^7\himsun$, for
the high-resolution region, and $\sim 2.5\times 10^5$ particles (whose
mass increase with the distance from the boundary of the
high-resolution region) for the coarse region. The simulations are
evolved by 5000 time steps with a comoving softening length $\eta \sim
2.5\kpc$.  At the end, each halo contains $N\sim 4 \times 10^5$
particles within their virial radius. The contamination of the coarse
particles, measured by the ratio of the mass of the coarse particles
within the virial radius to the total virial mass, is small, about
$10^{-2}$ for these galactic haloes. These haloes have been used by
Chen \& Jing (2002) for studying the angular momentum distribution in
dark haloes. Because it is difficult to resimulate halos when there
are more massive close neighbours, the haloes selected here are the 20
percent most isolated haloes (other authors also used similar
selections when resimulating haloes of a galactic mass). This
selection however does not guarantee that the halo sample is complete and
typical. To make sure that our results derived from this sample are
valid for haloes in general, we will use the GIF simulations
(Kauffmann et al. 1999) to verify our conclusions.

We use the spherical overdensity algorithm (hereafter SO) of Lacey \&
Cole (1994) to identify groups in the halo simulations. This algorithm
selects spherical regions whose average density is equal to
$\Delta_{\rm vir}$ times the mean cosmic density ${\bar \rho}$. Here
we adopt the fitting formula
$\Delta_{\rm vir}=(18\pi^2+82x-39x^2)/\Omega(z)$ of Bryan \& Norman (1998)
for $\Delta_{\rm vir}$, where $x \equiv \Omega(z)-1$ and $\Omega(z)$
is the cosmic mass density parameter at redshift $z$. The value of
$\Delta_{\rm vir}$ ranges from $\sim 180$ at high redshift to $\sim
340$ at the present for our $\Lambda$CDM cosmology. We selected groups at a
total of 20 outputs for each simulation. The outputs are
logarithmically spaced in the cosmic scale factor $a$ from $z=13.6$ to
$z=0$.

Because we will study how a halo grows with time, we need to construct
the main branch of the merging tree for each halo.  Given a group 
of dark matter particles at a given output time (which 
we refer to as group 2), we trace all its particles back to an earlier 
output time. A group (group 1) at the earlier output is selected as the ``main
progenitor''of group 2 if it contributes the largest number of 
particles to group 2 among all groups at this earlier output.  
We found that more than half of the particles of group 1 
is contained in group 2. We refer to group 2 as the 
``main offspring'' of group 1. For each of the five
haloes, we use this method to construct the main branch of the merging
tree from $z=0$ to $z=13.6$.

\begin{figure}
\begin{center}
 \leavevmode\epsfxsize=16.0cm \epsfbox{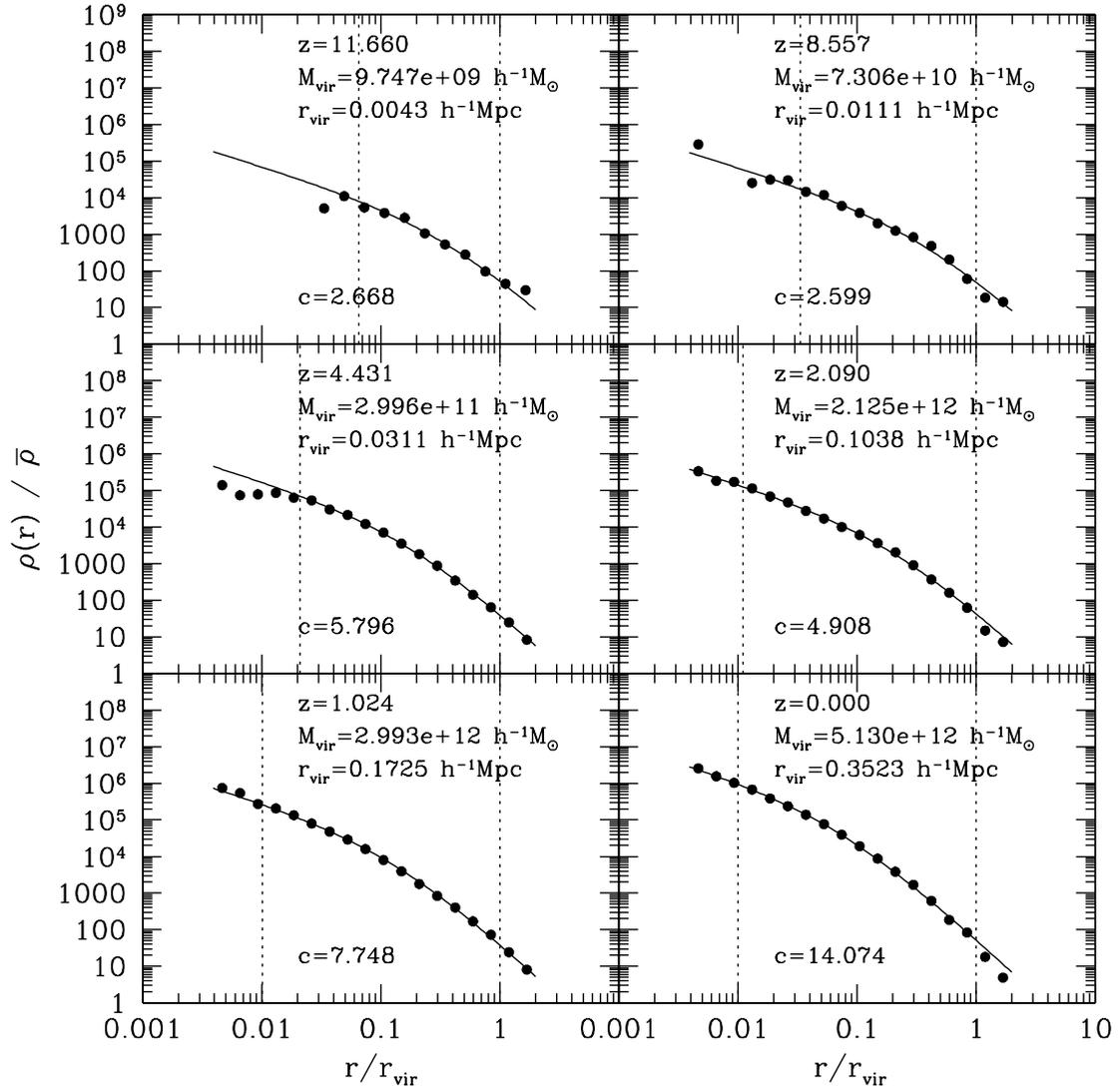}
\end{center} 

\caption{The density profiles of the main progenitors of halo 2
(symbols) at six output redshifts. The smooth curves are the NFW
fits to the profiles in the range $\eta < r < r_{\rm vir}$ 
(marked by the two vertical lines), where $\eta$ is the
force-softening length and $r_{\rm vir}$ is the virial radius. Also
shown in each panel are the concentration $c$, the mass $M_{\rm
vir}$ and the radius $r_{vir}$ of the main progenitors
at the redshift in consideration.
\label{fig:dpro}}

\end{figure}

The center of each progenitor group is chosen to be the particle
within the group which possesses the most negative gravitational
potential energy. The potential is calculated using only the group
particles.  We then inflate a sphere around this center until the
enclosed density drops below $\Delta_{\rm vir} {\bar \rho}$. The
selected region is called a virialized halo, and the
corresponding radius and the mass it encloses define $r_{\rm vir}$ and
$M_{\rm vir}$ for the halo. When constructing the halo density
profile, we include not only the particles assigned to the group but
also other surrounding particles. Then we fit the density profile of
each halo using the NFW form (Eq. \ref{eq:nfw}) and determine the best
fitting value of $r_s$. The radial bins are logarithmically spaced
from $\eta$ to $r_{\rm vir}$. If any bin includes less than 5
particles we decrease the number of bins by one until this is no
longer the case. The fitted value of $r_s$ is insensitive to the
choice of $r_{\rm vir}$ for the outer boundary.  As an example, we 
show in Fig. \ref{fig:dpro} the density profiles and the
corresponding NFW fit results of the main progenitors 
of one (the second) of the five haloes. 
 
In the literature, the concentration parameter $c$ is often used to
characterize the shape of the density profile, and is defined as the
ratio of the outer radius of the halo $r_h$ to $r_s$
(cf. equation \ref{eq:rhosc}). The definition for $r_h$ is still quite
arbitrary in the community; some authors opt to use $\Delta_h=200$
(e.g. Jenkins et al 2001) or $\Delta_h = \Delta_{\rm vir}$
(e.g. B01; Jing \& Suto 2002), while others choose the
halo mean overdensity to be 200 times the critical density of the
universe, i.e. $\Delta_h=200/\Omega(z)$ (NFW). These different
definitions of $r_h$ can lead to sizable difference in $c$ for a
given halo, and the difference depends on cosmological
parameters.  Although the difference in $r_h$ for the different
definitions is only a constant factor and can be corrected
(see Jing \& Suto 2002), we think it better to use $r_s$ 
to characterize the inner halo profile. In our following 
discussion, we will use all three definitions when
showing results for $c$.

\section{Mass accretion history and halo structure}

In Fig. \ref{fig:mass}, we show the mass-accretion histories, i.e. the
increase in mass of the main progenitors, for the five haloes. As one
can see, the build-up of a present-day galactic halo consists of an
early phase of fast accretion, where the halo mass $M_h$ increases
with time much faster than $H^{-1}(z)$, and a late phase of slow
accretion, where $M_h$ increases with time approximately as
$H^{-1}(z)$. The redshift $z_{tp}$ (called the turning point in the
following) which separates these two phases is roughly 3 for the
haloes considered here. On average, major mergers dominate the halo
formation at high redshift, and minor mergers dominate the slow
accretion phase. For the 5 haloes considered here, about 4/5
of the haloes undergo a major merger (with mass ratio larger than
$1/3$) within a Hubble time in the fast accretion phase, while this
fraction decreases to 1/4 in the slow accretion phase. In
Figs. 2 -- 6 we also show the evolution histories for other halo
properties (inner mass $M_s$, halo radius $r_h$, inner radius $r_s$,
halo circular velocity $v_h$, inner circular velocity $v_s$, inner
density $\rho_s$ and concentration $c$).  These figures show that
haloes in the two accretion phases show systematically different
properties.  For example, in the slow accretion phase, both the inner
scale radius ($r_s$) and the density at it ($\rho_s$) change slowly,
while in the fast accretion phase $r_s$ increases, and $\rho_s$
decreases, rapidly with the Hubble expansion. In the slow accretion
phase, $v_h$ and $v_s$ increase slowly and $v_s$ is higher than $v_h$,
while in the fast accretion phase, both $v_h$ and $v_s$ increase
rapidly with the Hubble expansion, and $v_h$ is slightly higher than
$v_s$. The halo concentration increases rapidly with the Hubble
expansion in the slow accretion phase, but has a slower change in the
fast accretion phase.  For haloes 1 and 4, where the early mass
accretion is much faster than $H^{-1}(z)$, $c$ is almost a constant.
   
 For an NFW halo with concentration $c$, the mean specific
binding energy is 
\begin{equation}
\label{E_binding}
{\cal E} =-{v_h^2\over 2} f_c, 
~~~~
f_c\approx {2\over 3}+\left({c\over 21.5}\right)^{0.7}
\end{equation} 
(see Mo, Mao \& White 1998).
If $c$ changes from 5 to 15 (the typical range for CDM haloes
in the slow accretion phase), $f_c$ changes from $\sim 1$
to $\sim 1.4$. Thus, in the slow accretion phase where $v_h$ is
almost a constant, the specific binding energy changes 
only slowly as $c$ increases, while in the fast accretion 
phase the potential well builds up rapidly as $v_h$ increases.
This suggests that the potential well associated with a halo
is built up mainly during the early phase of rapid 
accretion. Note that for the five haloes considered 
here, the total mass increases by a factor of 3 to 10
while $v_h$ changes only a little during the slow 
accretion phase. Apparently, such slow accretion does not 
affect much the potential well, although it can cause
a large increase in halo mass.

 From equation (\ref{E_binding}) we see that for $c\approx 5$ (so that
$f_c\sim 1$) the mean specific binding energy of an NFW halo is the
same as that of a singular isothermal sphere with the same circular
velocity. For an NFW halo, this occurs at a time when $v_h\approx v_s$
($c_{\rm 0}=4.86$ for $v_h = v_s$). As one can see from Figs. 5 and 4,
this time follows closely the epoch when the accretion makes the
transition from the fast to the slow phase. This transition defines a
characteristic time for the mass-accretion history of a halo, as we
will see in the following.

\section {Predicting halo structural properties from 
mass-accretion histories}

As one of the main results of this work, we find that the 
scale radius $r_s$ is tightly correlated with the scale mass $M_s$ 
(as shown in Fig. \ref{fig:msrs}), 
which can be well represented by a simple power law:
\begin{equation}
\label{eq:msms0}
   {M_s \over M_{s,0}}=f\left({r_s\over r_{s,0}}\right), 
~~~~\mbox{with}~~~~~ f(x)=x^{3\alpha}\,,
\end{equation}
where $M_{s,0}$ and $r_{s,0}$ are the scale mass and scale radius
at some chosen epoch, say the turning point $z_{tp}$. The slope $\alpha$ 
remains about $0.6$ during the whole evolution histories 
of the haloes. More precisely, we find that the slope is slightly 
different in the two phases separated by $z_{tp}$. 
For the slow accretion phase ($z<z_{tp}$), 
$\alpha=0.48\pm 0.03$, while for the rapid accretion phase
($z> z_{tp}$), $\alpha=0.65\pm 0.04$ (see Fig. \ref{fig:msrs}).
During the evolution of a halo, the mass $M_s$ and $M_h$ 
have typically grown by a factor of about 3 and 10 respectively 
from $z_{tp}$ to $z=0$, and more than 1000 times from the
first output to $z=0$.  Over the entire redshift span, both 
the initial power spectrum at the mass scale of 
the progenitors and the cosmological parameters
[$\Omega(z)$ and $\lambda(z)$] change substantially, and so
the tight correlation between $M_s$ and $r_s$ we find
should be valid quite universally for CDM-class models.   
Further test of this scaling relation will be made below
by comparing its predictions with the results obtained 
from another set of $N$-body simulations.

An immediate application of the scaling relation is to use
it to predict the evolution of the density profile of a
halo from its mass-accretion history. This can be done as follows.
For the NFW profile we have 
\begin{equation}
\label{eq:mhms}
   {M_h \over M_s}=g\left({r_h\over r_s}\right), \;\;\;\; {\rm with}
   \;\;\;\; g(x)={{\ln(1+x)-x/(1+x)}\over \ln 2-1/2}.
\end{equation}
Combining equations (\ref{eq:msms0}) and (\ref{eq:mhms}), we get
\begin{equation}
\label{eq:mhms0}
   {M_h\over M_{s,0}}=f\left({r_s\over r_{s,0}}\right)g\left({r_h\over
   r_s}\right).
\end{equation}
Thus, for a given halo definition, which relates 
$M_h$ and $r_h$, and a given calibration, which specifies
$r_{s,0}$ and $M_{s,0}$, the scale radius
$r_s$ is uniquely determined by the halo mass $M_h$
through the above equation. We can therefore obtain
$r_s$ as a function of time, once the mass-accretion history 
is known. Equivalently, in terms of $c$ we have
\begin{equation}
\label{eq:cmh}
   {[\ln(1+c)-c/(1+c)]c^{-3\alpha} \over
   [\ln(1+c_0)-c_0/(1+c_0)]c_0^{-3\alpha}}=\left[{\rho_h(z)\over
   \rho_{h,0}}\right]^\alpha\left[{M_h(z)\over
   M_{h,0}}\right]^{1-\alpha}\,.
\end{equation}
Again, once a calibration is adopted to specify $c_0$ and $M_{h,0}$,
we can predict $c$ as a function of redshift for a given
mass-accretion history $M_h(z)$. From this equation, one can see that
if the increase of the second term dominates the decrease of the first
term on the right hand side, i.e. the log mass accretion rate is
higher than $H^2(z)$ for $\alpha=0.5$ (as in the fast accretion
phase), the predicted $c$ will decrease rather than increase.

The existence of the tight relation between $M_s$ and $r_s$ means that
the structural properties of a halo can be related to its
mass-accretion history.  The scaling relation obtained here is
different from that assumed in W02, $c\propto a$.  For large $c$,
equation (\ref{eq:cmh}) gives $c\propto (1+z)^{-1}
M_h^{(1-\alpha)/3\alpha}$, which is the same as the scaling proposed
by W02 only if $\alpha=1$. Since $\alpha\sim 0.5$, equation
(\ref{eq:cmh}) implies that $c$ also depends on the mass accretion
[i.e. on the form of $M_h(z)$].  As we will show later in this
section, our scaling relation leads to more accurate predictions for
halo structural properties than the assumption of W02 does.

In order to use equation (\ref{eq:mhms0}) or (\ref{eq:cmh}) to
predict $r_s(z)$ and $c(z)$, one has to specify the concentration $c$
(or the scale radius $r_s$) at some fiducial time (i.e. to calibrate
the zero-point of the scaling relation).  If the mass-accretion
history of a halo is given by an $N$-body simulation, the zero-point
calibration can be made at the final output time of the simulation,
and our formulae can be used to predict the halo concentration along
the main branch of the merging tree at any other time.  Since the
determination of the mass-accretion histories of dark haloes does not
require the resolution of density profiles of small objects at high
redshift, our model allows one to assign concentration parameters to
high-redshift haloes even in simulations with moderate resolutions.

\begin{figure}
\begin{center}
 \leavevmode\epsfxsize=16.0cm \epsfbox{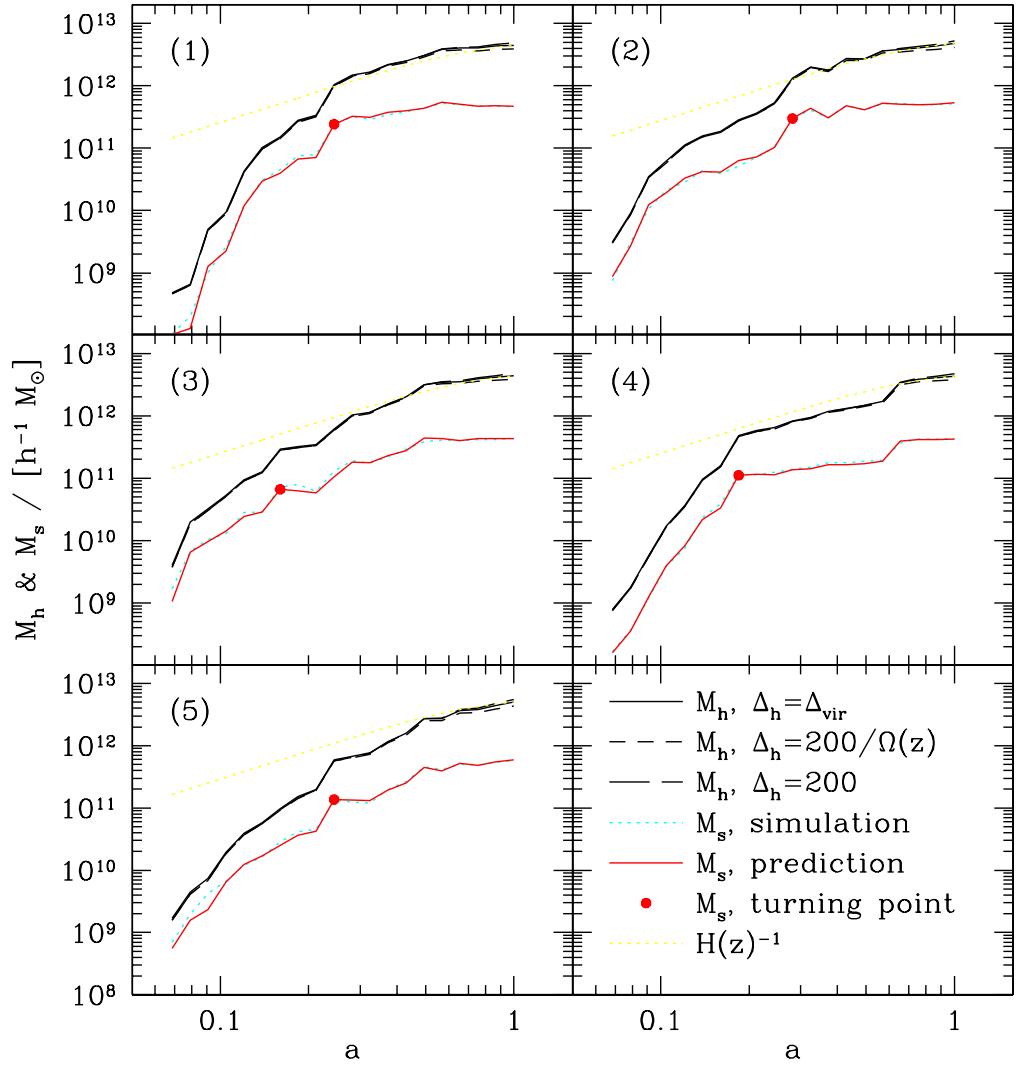}
\end{center} 
\caption{The mass accretion histories (i.e. the change of the mass of
the main progenitor with the scale factor) for the five
high-resolution haloes. The halo mass $M_h$ depends on the
overdensities $\Delta_h$ used to define a halo, but the difference
among the three definitions we have used is quite small and almost
invisible in the figure. In each panel we also show (as the lower
dotted curve) the evolution of the mass $M_s$ within the scale radius
$r_s$ measured directly from the simulations.  The lower solid curves
show the predictions for $M_s$ by the model proposed in the text.
Note that the model prediction for $M_s$ follows very closely the
simulation results. The filled circle in each panel marks the turning
point (in the mass-accretion history) obtained using the model
described in Section 4. The upper dotted curves show the scaling
relation $M_h \propto H(z)^{-1}$ which follows the mass accretion
$M_h(z)$ in the slow accretion phase.
\label{fig:mass} }
\end{figure}

There is another calibration scheme which can be implemented in
semi-analytical models of halo merger histories (obtained via, e.g.,
the extended Press-Schechter theory).  As shown in
Fig. \ref{fig:mass}, the growth rate of halo mass changes dramatically
around $z_{tp}$.  This epoch marks a turning point in the mass
accretion history, and may be chosen to calibrate the zero point. We
will propose an objective recipe to find out the characteristic time.
As shown in Figs. \ref{fig:mass} and \ref{fig:velocity}, even though
the mass-accretion history differs substantially from halo to halo,
all haloes have similar asymptotic behavior: $M_h(z) \propto
H^{-4}(z)$ and $v_h(z) \propto H^{-1}(z)$ at the early fast accretion
phase, and $M_h(z) \propto H^{-1}(z)$ and $v_h(z) \sim$ constant in
the late slow accretion phase. After some trials, we found a good
approximation to the characteristic time ($z_{tp}$) which is given by
the epoch when $\log (v_h(z))-\gamma\log (H(z))$ reaches its maximum,
where $\gamma$ is a constant.  We will use $\gamma=-1/4$, but our
results are robust to the change of $\gamma$ from $-1/8$ to
$-1/2$\footnote{We have tested this by changing $\gamma$ from $-1/2$
to $-1/8$. There are only small systematic changes ($\sim 10\%$) in
the results which can be eliminated by adjusting the value of $c_0$,
the reason may be that the turning is sharp.}. The typical
concentration at the turning point is $c=4.0$.  The zero-point
calibration can then be made by setting $c_{\rm 0}=4.0$ and $M_{h,0}$
equal to the mass of the main progenitor at $z_{tp}$. This calibration
method is used for prediction throughout the paper.

The basic idea in the above calibration of $c$ is the same as that of
W02, based on the simulation results that the halo concentration at
the end of a fast accretion phase is approximately at a constant value.
In the recipe proposed by W02, the time at which mass accretion changes
from the fast phase to the slow phase is obtained by fitting the mass
accretion history by an exponential form. Here we do not assume any
universal form for the mass accretion history.
 
\begin{figure}
\begin{center}
 \leavevmode\epsfxsize=16.0cm \epsfbox{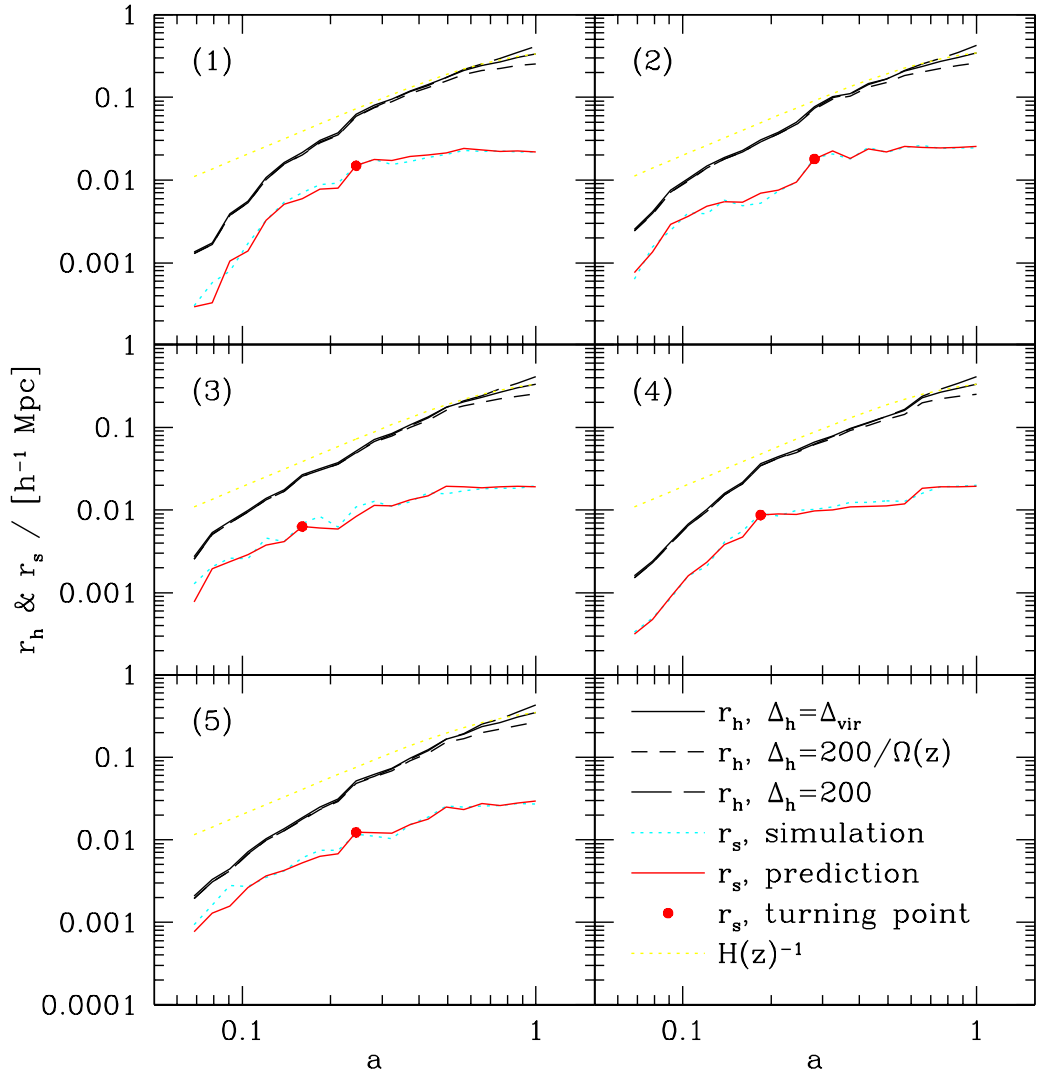}
\end{center} 
\caption{The evolutions of halo radius
$r_h$ and inner radius $r_s$. The notations 
are the same as in Fig.\ref{fig:mass}. 
\label{fig:radius} }
\end{figure}

\begin{figure}
\begin{center}
 \leavevmode\epsfxsize=16.0cm \epsfbox{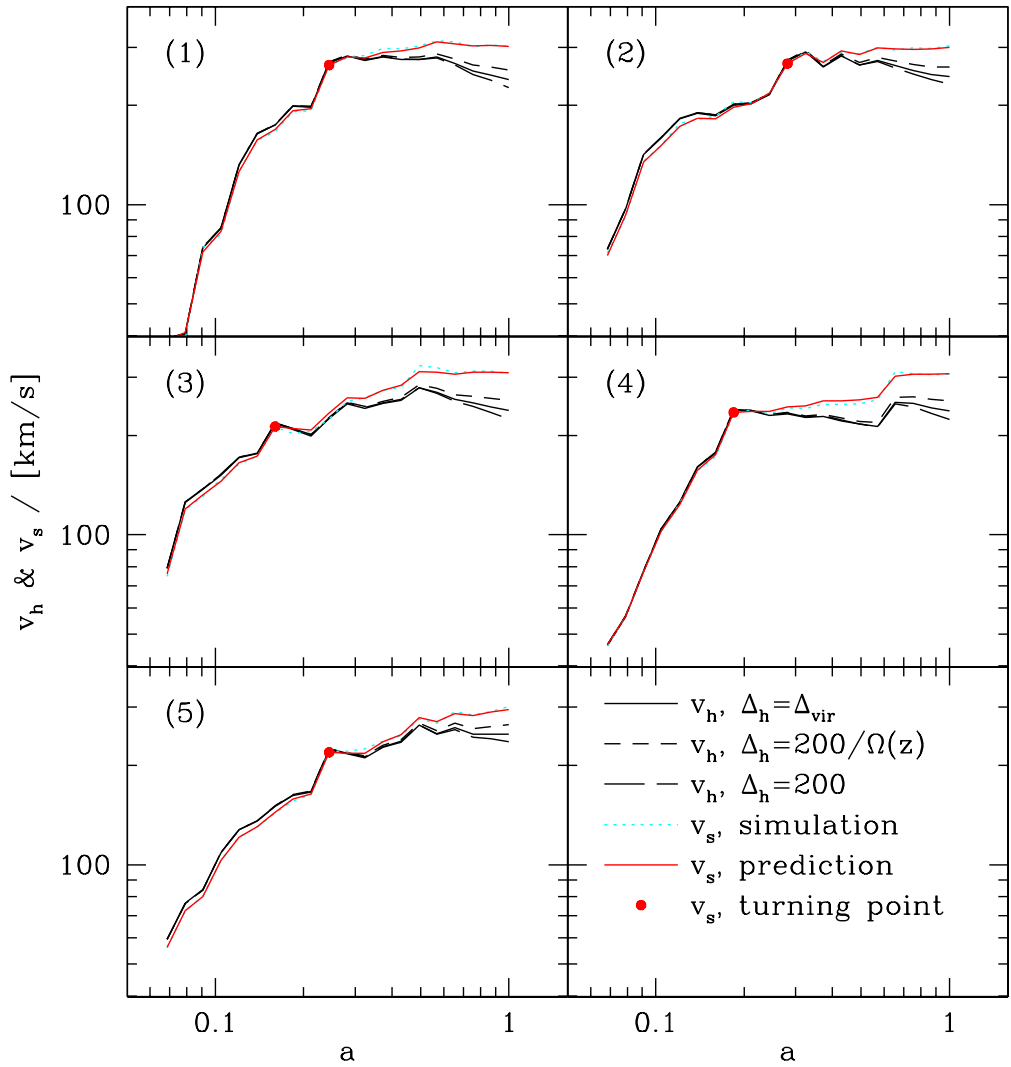}
\end{center} 
\caption{The evolutions of halo circular velocity
$V_h$ and inner circular velocity $V_s$. The notations 
are the same as in Fig.\ref{fig:mass}. 
\label{fig:velocity}}
\end{figure}

\begin{figure}
\begin{center}
 \leavevmode\epsfxsize=16.0cm \epsfbox{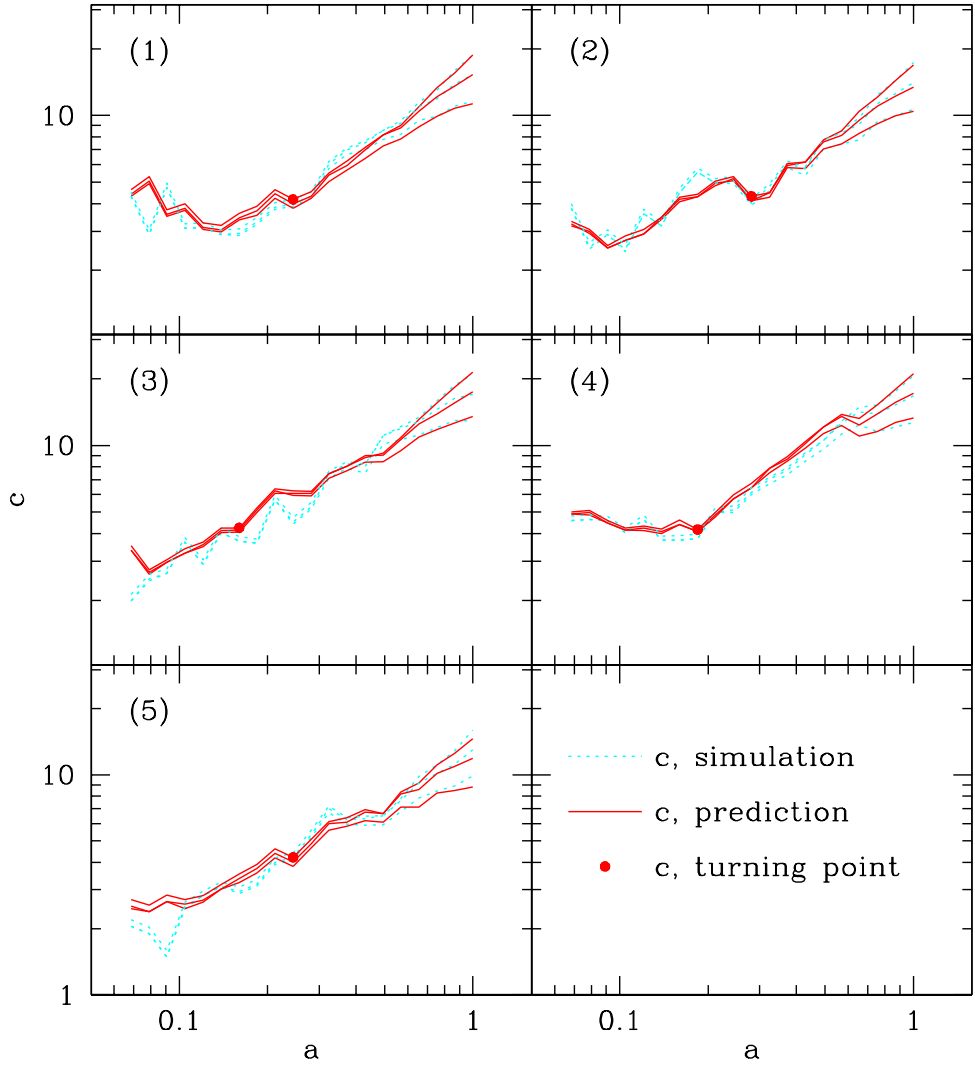}
\end{center} 
\caption{Halo concentration parameters $c$ measured 
from the simulations (dotted lines), 
compared with the model predictions based on the recipe
described in the text (solid lines). 
The concentration depends on the halo definition, 
and the three lines of a given type in each panel 
correspond, from bottom up, 
to $\Delta_h=200/\Omega(z)$, $\Delta_{\rm vir}$, and
$200$, respectively. The filled circle in each panel 
marks the turning point in the mass-accretion history.
\label{fig:cc} }
\end{figure}

\begin{figure}
\begin{center}
 \leavevmode\epsfxsize=16.0cm \epsfbox{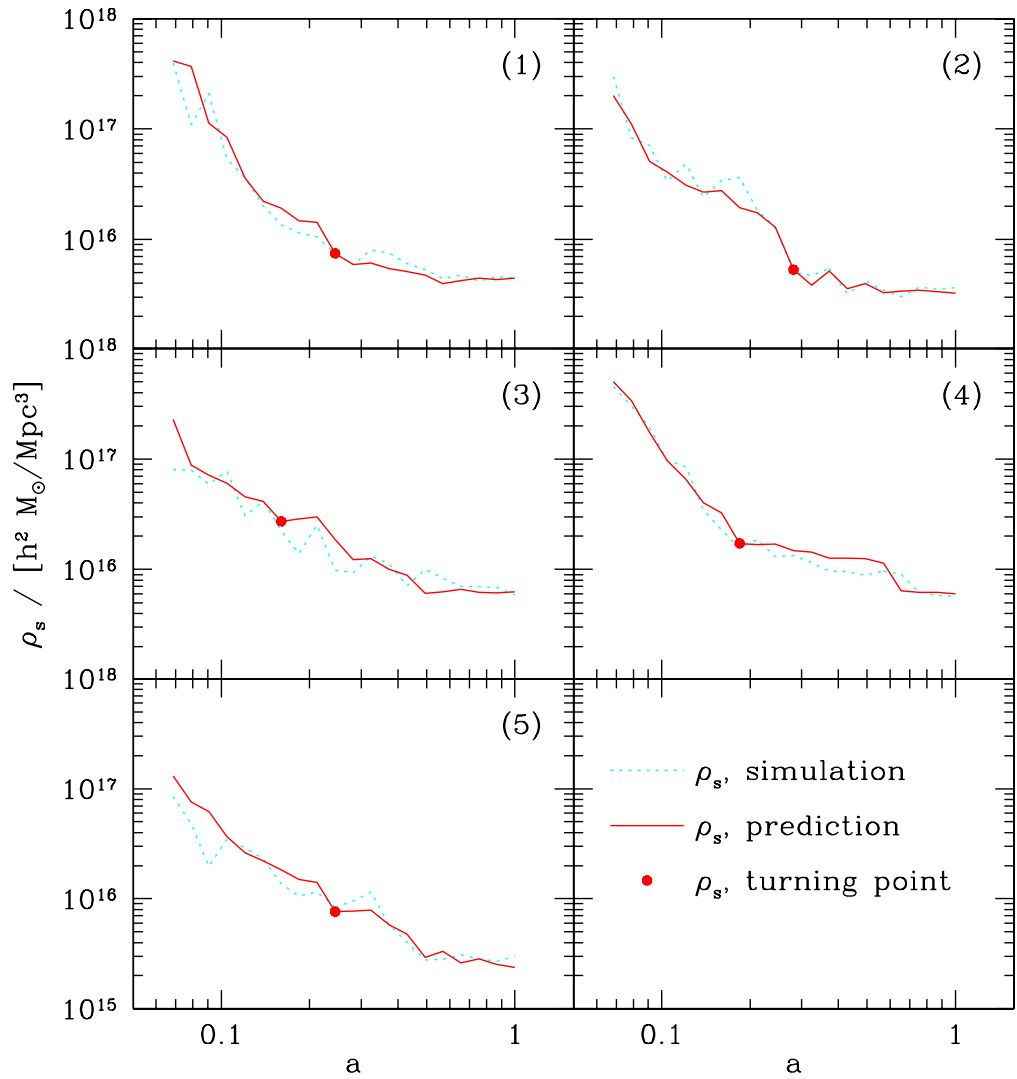}
\end{center} 
\caption{The evolution of the density ($\rho_s$) 
at the scale radius ($r_s$) given by the simulations 
(dotted lines), compared with model predictions
(solid lines). The filled circle in each panel marks the 
turning point of the mass-accretion history.  
\label{fig:density} }
\end{figure}

\begin{figure}
\begin{center}
 \leavevmode\epsfxsize=16.0cm \epsfbox{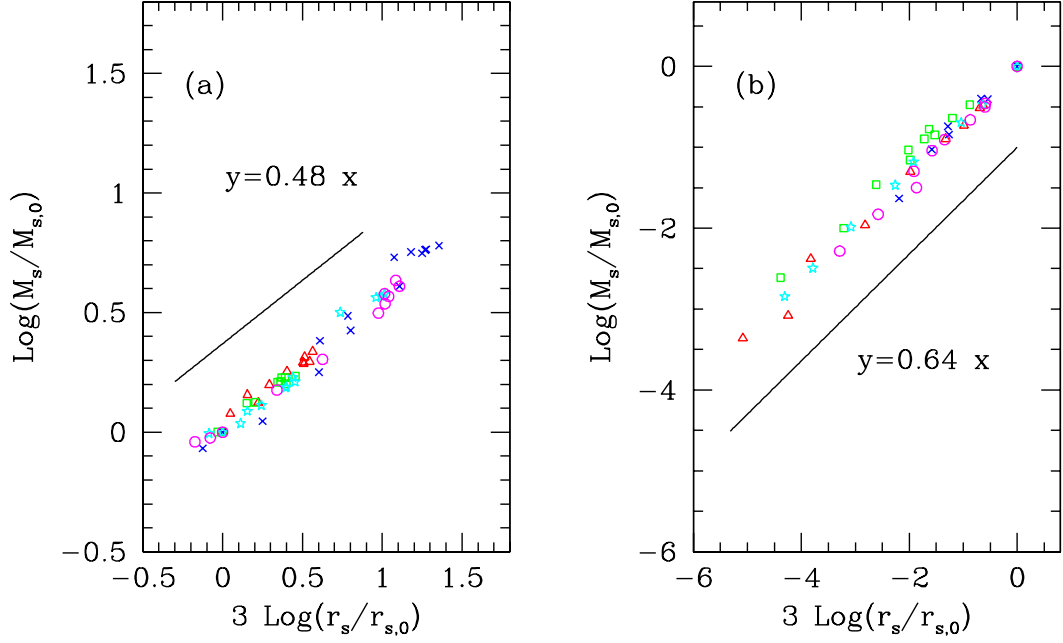}
\end{center} 
\caption{The correlation between the inner mass $M_s$ (i.e. the mass
enclosed in the scale radius $r_s$) and $r_s$ for haloes at different
output times.  These two quantities are scaled by $M_{s,0}$ and
$r_{s,0}$, the values at the turning point of the mass accretion
history. Different symbols are used for different haloes, and the
results support that there exists a well-defined scaling relation
between $M_s$ and $r_s$.
\label{fig:msrs} }
\end{figure}

Once the mass-accretion history is known, the calibrated 
relation (\ref{eq:cmh}) can be used to predict the structural
quantities of a halo, such as $m_s$, $r_s$, $v_s$, $c$, 
and $\rho_s$. The predictions for these quantities are 
shown in Figs. \ref{fig:mass} -- \ref{fig:density}.  
We have used $\alpha=0.48$ for $z<z_{tp}$
and $\alpha=0.65$ for $z>z_{tp}$, as derived from
the simulations. The predictions are compared 
with quantities measured directly from the simulations. As one can
see, our recipe works pretty well even when the halo mass $M_h$ 
and the inner mass $M_s$ grow by factors of several hundreds.

Since our model is based only on simulations 
of five present-day galaxy haloes, it is necessary
to test its validity against other simulations.
Here we use the GIF, a set of cosmological N-body 
simulations carried out at the MPA (e.g. Kauffmann et al. 1999).
The GIF simulation we use here was performed with 
$N=256^3$ particles in a comoving $141.3\mpc$ box, and so the
particle mass is $m_p \sim 1.4\times 10^{10} \himsun$. The comoving
softening length $\eta \sim 20 \kpc$, and the cosmogony is the same as
that of our halo simulations except that $\sigma_8=0.9$. 
Dark matter particles are grouped using the standard 
friend-of-friend algorithm with a linking length  
0.2 times the mean separation of particles. The halo mass
$M_{200}$ within $r_{200}$ and the merger tree information at 44
output times (logarithmically spaced in $a$ from $z=12.2$ to $z=0$)
are provided . We use this tree information to trace the main
progenitors of a halo backwards step by step to $z=5.80$, re-locate the halo
center to the most bound particle, and calculate the halo mass using the
three halo definitions discussed in Section 2. 

\begin{figure}
\begin{center}
 \leavevmode\epsfxsize=16.0cm \epsfbox{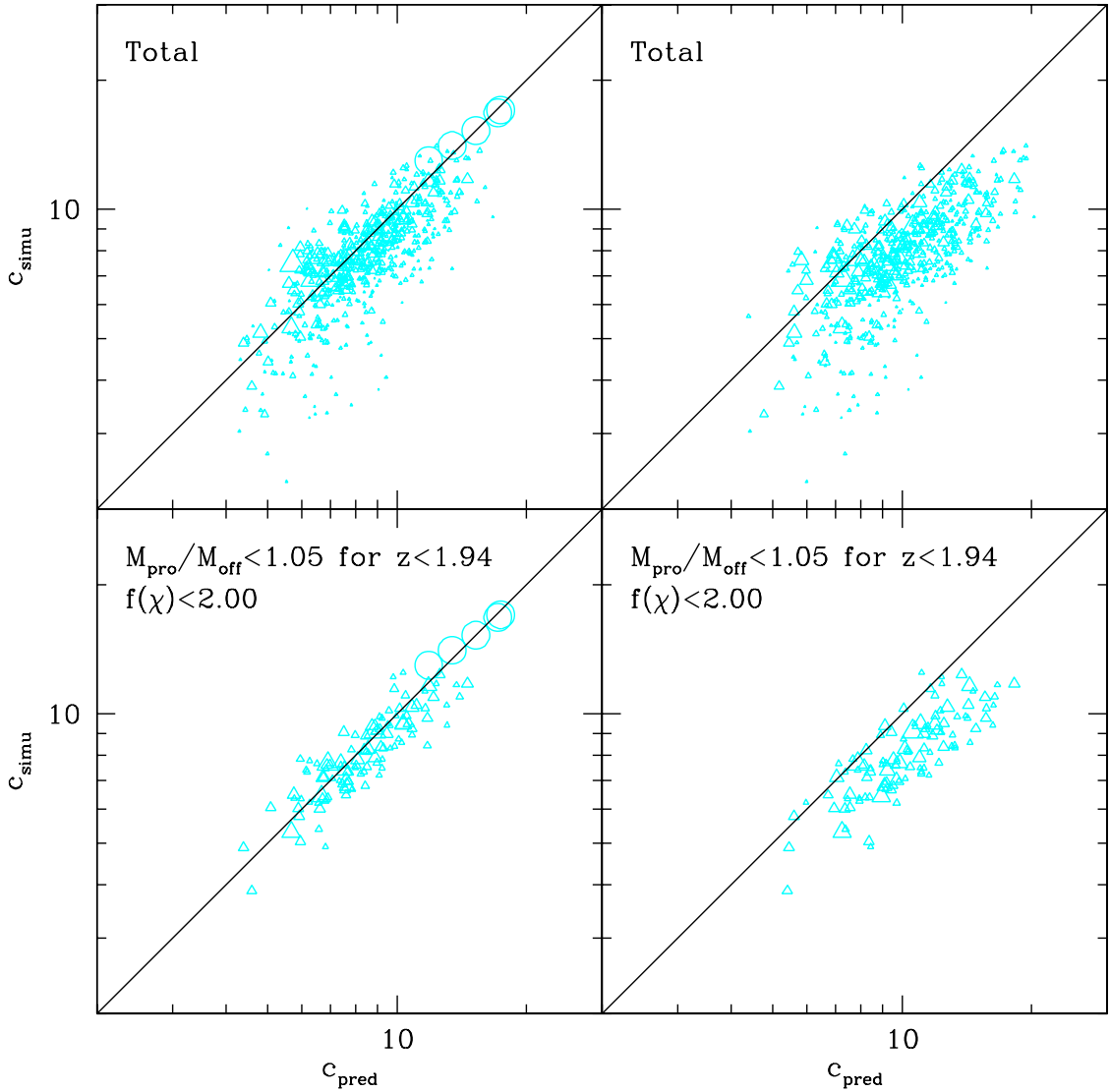}
\end{center} 
\caption{Left panels -- The concentrations measured from the GIF
simulation versus model predictions based on our recipe
(see text for details). Note that the GIF haloes are more massive 
than $2.2 \times 10^{13} \himsun$ at $z=0$ (triangles). 
The results for the five high-resolution galaxy haloes 
(circles) are also plotted for comparison.  Here haloes are 
identified with the definition $\Delta_h=\Delta_{\rm vir}$. 
The size of a symbol is inversely proportional to the NFW fitting 
error. The lower left panel is the same as the upper left panel, 
but for haloes with better quality in the NFW fitting 
($f(\chi)<2$) and smaller error in their mass 
accretion history ($M_{\rm pro}/M_{\rm off}<1.05$). 
Right panels -- the same as the left panels, but the 
predictions are made using the model of
Wechsler et al. (2002). \label{fig:gif1}}
\end{figure}

\begin{figure}
\begin{center}
 \leavevmode\epsfxsize=16.0cm \epsfbox{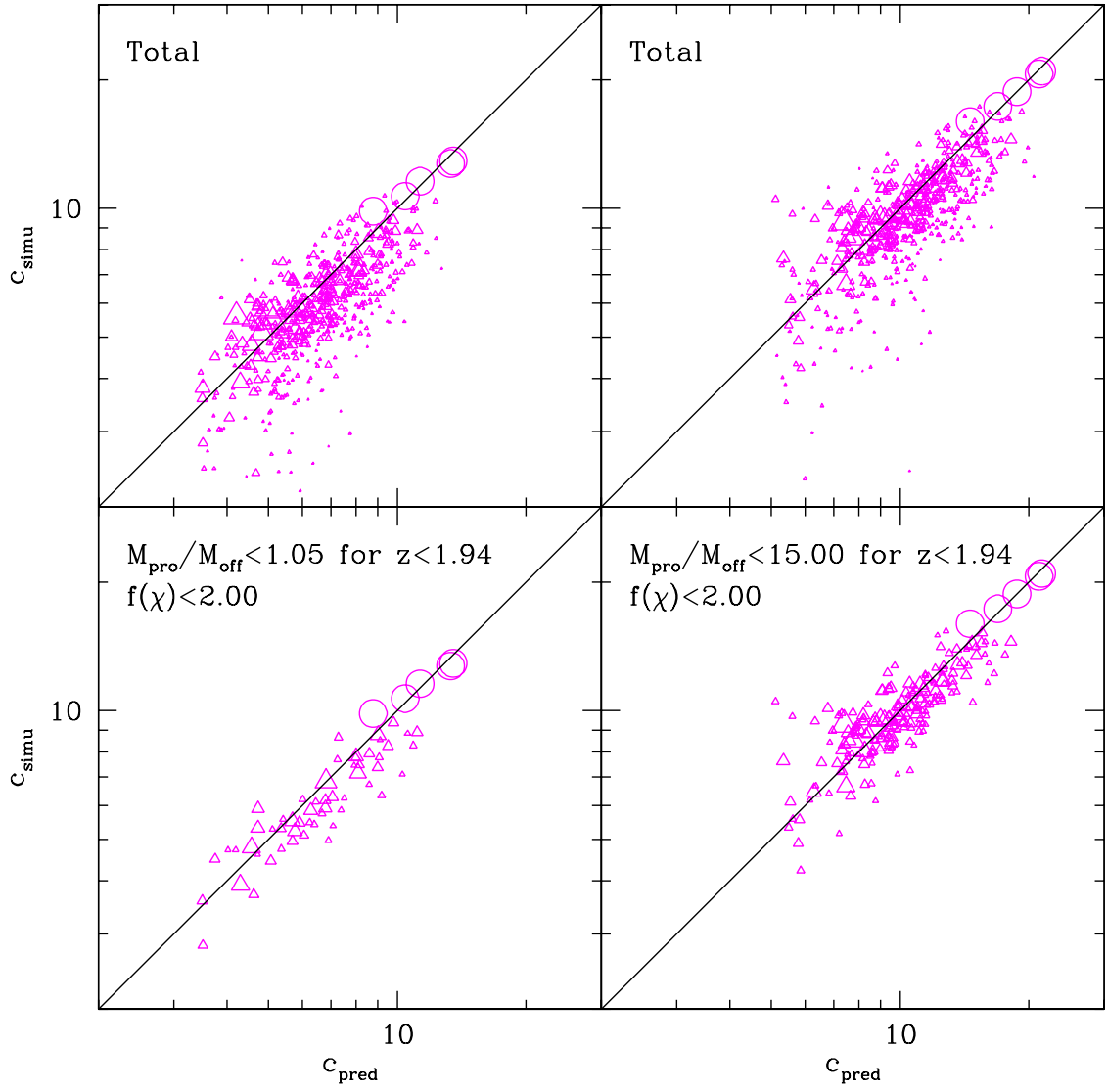}
\end{center} 
\caption{The same as left panels of Fig.~\ref{fig:gif1}, except that 
haloes are defined using $\Delta_h=200/\Omega(z)$ (left panels) and 
$\Delta_h=200$ (right panels).
\label{fig:gif2} }
\end{figure}

We predict the final halo concentrations for haloes more massive than
$2.2\times 10^{13} \himsun$ (i.e. those containing more than $1600$
particles) using our recipe, and compare them with the results of
direct NFW fitting to the simulation data. Our recipe is not
applicable to a small fraction ($\sim 3\%$) of haloes for which the
mass growth has always been fast and so $z_{tp}$ cannot be
determined. Note that even for these cases a calibration can still be
made in semi-analytical models by extending the merger tree further in
time. For the five haloes considered above, $z_{tp}\gg 1$ [and so
$\Omega(z_{tp})\approx 1$ and $\lambda(z_{tp})\approx 0$] and so all
the three definitions of $\Delta_h$ give essentially the same
concentration, $c\approx4.0$ at $z_{tp}$. However, for the more
massive haloes in the GIF simulation, the values of $z_{tp}$ for some
haloes are close to zero and the value of $c$ depends on halo
definition.  We have tested the calibration for all the three
definitions, and found that the calibration $c_0=4.0$ at $z_{tp}$
holds best for the definition $\Delta_h=\Delta_{\rm vir}$.  For the
other two definitions, one has to adjust the value of $c_0$, but this
can be done straightforward since the difference of halo density among
different definitions is known at any given redshift\footnote{Actually
this has already been included in the prediction for the above 5
haloes}.  The predictions based on this calibration are compared with
the simulation results in the upper left panel of Fig. \ref{fig:gif1}
and upper panels of Fig. \ref{fig:gif2} for all the three
definitions. The scatter of the correlation between the model
predictions and the simulation results (i.e. around $c_{\rm
simu}=c_{\rm pred}$) is $22.4\%$, $26.5\%$ and $23.1\%$ respectively.
Several sources can cause this scatter: (i) the inaccuracy of the
scaling relation (Eq. \ref{eq:msms0}) and the error in its
calibration; (ii) the error of halo identification and mass counting
from simulation; and (iii) the error in the NFW fitting of halo
profiles due to the finite number of particles. To demonstrate these
effects, we show in the lower panels of these figures the results for
about $20\%$ of the haloes which have better qualities in NFW fitting
and better defined halo mass. The quality of NFW fitting is
represented by $f(\chi)$, which gives the relative fitting chi-square,
while the quality of a mass accretion history is defined as the
maximum of the ratio $M_{\rm pro}/M_{\rm off}$, where $M_{\rm pro}$
and $M_{\rm off}$ are, respectively, the masses of the progenitor and
offspring at any adjacent outputs. An accretion history with
progenitors much more massive than the offsprings is not well defined,
because accretion should increase, rather than decrease the halo mass
under normal circumstances. The scatter for the high-quality data is
$12.5\%$, $15.5\%$ and $13.6\%$ correspondingly. The reduction in the
scatters suggests that part of the scatters is due to the inaccuracy in
the NFW fitting and in the counting of halo mass.
   
It is worth repeating that the GIF haloes analyzed here have masses
larger than $ 2.2\times 10^{13} \himsun$, and are formed
recently. This means that with these haloes we are actually probing
quite different scales, formation times and cosmological models
[$\Omega(z)$ and $\lambda(z)$] from those in our high-resolution halo
simulations. A very good agreement of our predictions with the GIF
simulations reinforces the conclusion that our recipe is valid for
both galaxy and cluster haloes, independent of the halo definitions
and cosmological parameters. In a forthcoming paper (Zhao et
al. 2002, in preparation), these conclusions are confirmed by new
results from simulations with much higher mass resolution and in
different cosmogonies.

As pointed out in the introduction, W02 have used a similar recipe to
predict halo concentrations from mass-accretion histories.  For
comparison, we show the results using their method in the right panels
of Fig. \ref{fig:gif1}. The scatter around $c_{\rm simu}=c_{\rm pred}$
for the total sample and the $20\%$ high qaulity haloes is $40.3\%$
and $31.6\%$ respectively.  The predicted $c$ based on the W02's
recipe is about 20 percent higher $c$ than the simulation results. We
noted that the calibration of W02 was done for haloes of mass $2\times
10^{12}\msun$, while the haloes considered here are at least ten times
more massive. The systematic offset may be caused by the different
ranges of halo mass considered.  After submitting this paper, we have
analyzed several high-resolution simulations of $256^3$ particles in
different boxsizes and different cosmogonies (Zhao et al. 2002, in
preparation), where we not only have confirmed that the W02's prediction is
accurate for LCDM haloes of $2 \times 10^{12}\msun$ but also have found
that the W02's prediction is systematically higher for cluster haloes
but lower for dwarf galactic haloes. Therefore, simply adjusting the
calibration by multiplying a constant in the W02's recipe does not
work for haloes over a wide range of mass.  The reason might be that,
even in the late slow accretion phase, $c$ isn't simply proportional
to $a$ as W02 assumed. In fact, W02 also noticed that the increasing
rate of $c$ is different for different halo populations, but they
didn't include it in their model, since otherwise a series of
assumption about the increasing rate of $c$ should be made.

For the small fraction ($\sim 3\%$) of haloes where our calibration is
not applicable (as mentioned earlier in this section), an exponential
fit to the mass-accretion history can still return a value for the
formation time, and so the calibration of W02 is still
applicable. However, the formation times obtained for such cases are
very uncertain, because they are obtained by the extrapolation of a
small segment of the mass-accretion history.

\section{Discussion}

In this paper we have studied how the structural properties of 
CDM haloes are correlated with their mass-accretion histories.
We have found a tight scaling relation between the inner mass $M_s$
and the inner scale radius $r_s$ and used this scaling relation to build
a model to predict the structural properties of dark haloes
from their mass-accretion histories. Comparing the model predictions 
with results from $N$-body simulations show that our model is 
more accurate than related models proposed earlier based on simpler 
assumptions. Combined with halo merging trees constructed either from 
moderate resolution simulations or semi-analytical models,
our recipe allows one to trace the evolution of the structural 
properties of dark haloes, and so is useful for constructing
model realizations of galaxy formation in dark matter haloes.

 Our results also provide important insights into the formation
process of dark matter haloes in the CDM cosmogony, and so can help us
to understand the structural properties of dark matter haloes.  In
agreement with W02, the formation of a dark halo are found to consist
of an early phase of fast mass accretion, and a late phase of slow
mass accretion. We have found that this two phases are separated at a
time when the circular velocity at the inner radius is about equal to
the halo circular velocity at the virial radius. We have shown that
the formation process in these two phases have different
properties. In particular, the inner structure and the potential
associated with a halo are built up mainly in the fast accretion
phase, even through large amount of mass can be accreted to the system
in the slow accretion phase.  Above all, we found that $c$ isn't
uniquely proportional to $a$, and it grows more slowly when accretion
is faster. This is quite different from the W02 recipe's assumption.

\begin{figure}
\begin{center}
 \leavevmode\epsfxsize=16.0cm \epsfbox{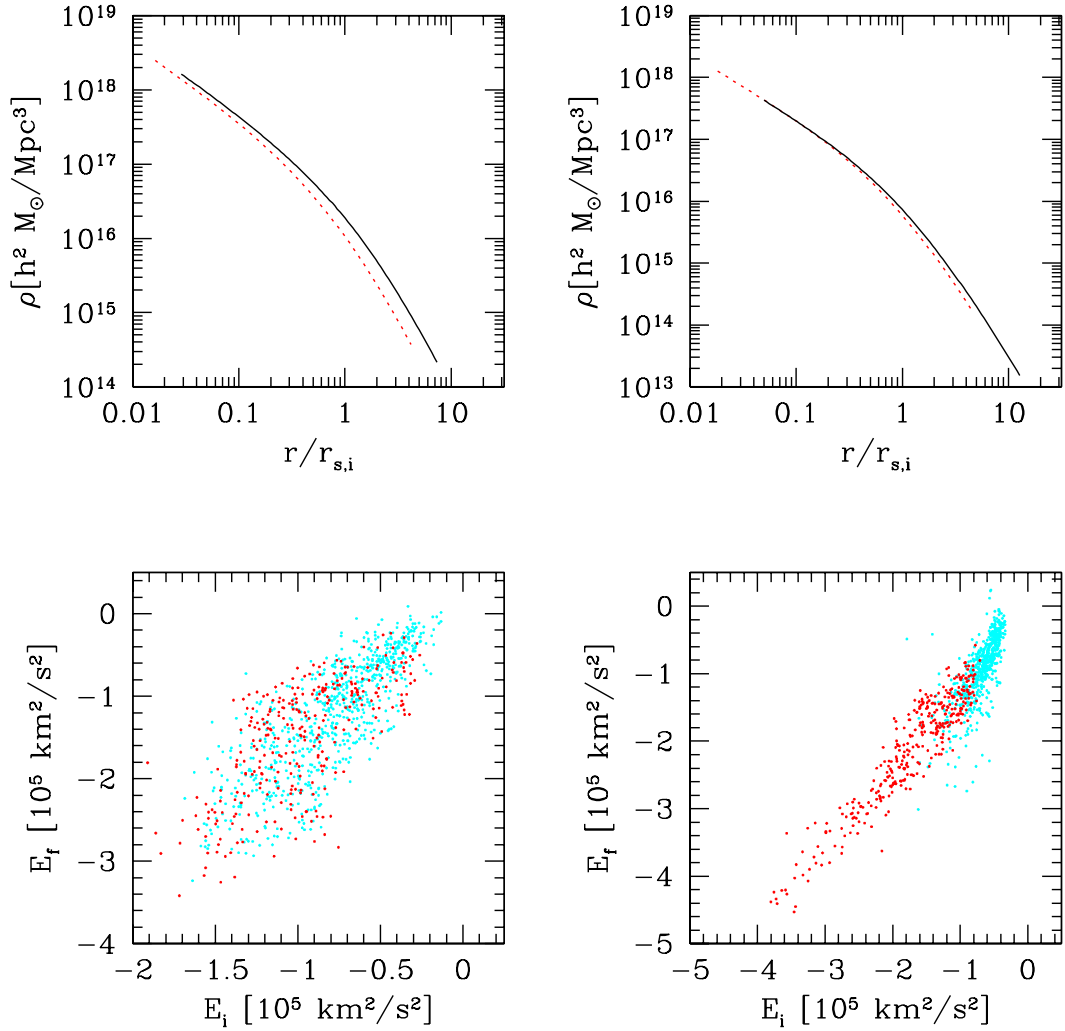}
\end{center} 
\caption{The changes in halo density profile (upper panels)
and particle specific energy (lower panels) during 
a time interval (in which the mass of the halo increases 
by a factor of about 3) in the fast (left panels) and 
slow (right panels) accretion phases. 
The dashed and solid lines in the upper panels 
are the initial and final halo profiles, respectively.
The radius is scaled with the initial inner radius, $r_{s,i}$. 
Lower panels show the final versus initial specific
binding energy of 1000 randomly selected particles.
Red dots are particles in the main progenitor 
while blue dots are newly accreted particles during 
the time interval. 
\label{fig:enden}}
\end{figure}

It has been long speculated that the universal density profile of dark
haloes may be due to violent relaxation in gravitational collapse
(e.g. Lyden-Bell 1967; White 1996). Our results show that although
violent relaxation may have played an important role in the formation
of a dark halo in the fast accretion phase, the density profile in the
slow mass-accretion phase cannot be attributed to this process.  This
is illustrated in Fig. \ref{fig:enden}, where we plot, for both the
fast and slow accretion phases, the change in particle specific
binding energy and the halo density structure over a time interval in
which the halo mass increases by a factor of 3.  As one can see, there
is a tight correlation between the initial and final specific binding
energies in the slow accretion phase. Furthermore, in the slow
accretion phase, the accreted material retains its low binding energy
($E\sim 0$), and is added in the outer part of the halo without
altering much the internal structure (see the upper right panel). In
contrast, in the fast accretion phase, the accreted material is well
mixed with the main progenitor in energy space (the low left panel),
and the inner halo profile is significantly altered (the upper left
panel). Note that even in the fast accretion phase which involves
major mergers, there is still a significant correlation between the
initial and final binding energies, suggesting that the violent
relaxation is incomplete (see also Quinn \& Zurek 1988; Zaroubi \&
Hoffman 1993). These results suggest that the universal density
profile found for CDM haloes in $N$-body simulations is not completely
due to violent relaxation. Although violent relaxation may be
responsible for the build-up of the inner structure of a halo, gentle
accretion due to secondary infall also gives rise to an outer profile
which matches the universal form.  The reason for this is still
unclear, but our results suggest that the universal profile should be
understood at least in terms of violent relaxation and slow secondary
infall.

 Since in the standard model of structure formation, galaxies
are assumed to form by gas cooling and condensation in dark
matter haloes, our results also have important implications
for galaxy formation. For a given halo, the free-fall time
$r_h/v_h$ is roughly proportional to $H^{-1}(z)$. Thus, in the 
fast accretion phase where the mass accretion rate is faster 
than the free-fall time, dark haloes can hardly establish 
dynamical equilibrium before another major merger occurs. 
If the merger progenitors contain gas, the cold gas 
component will not be able to settle into an equilibrium 
thin disk in the fast accretion phase, but rather is 
constantly disturbed and compressed by the merger. 
The situation resembles closely that in merger-driven starbursts 
observed in the local universe (see Sanders \& Mirabel 1996
and references therein). Thus, associated with the early 
fast accretion phase may be an episode of rigorous star formation, 
which may be responsible for the formation of stars in galaxy bulges.
The formation of galaxy disks may then be associated with the late 
slow accretion phase. The state of the gas in a protogalactic region
may be significantly affected by the early star formation, and so the 
subsequent disk formation by secondary infall is expected to 
be influenced by the formation of bulge stars  
(see Mo \& Mao 2002 for a detailed discussion).   

  As we have seen, the halo concentration parameter
increases significantly with time in the slow accretion
phase, and so haloes at high redshifts are expected to be less 
concentrated. This may have important consequence for the formation 
of disk galaxies in dark haloes. As shown in Mo, Mao \& White (1998),
for a galaxy disk formed in a NFW halo, the central surface
density of the disk can be written as 
$\Sigma_0\propto H(z) V_h f_c/(\lambda f_R)^2$, 
where $V_h$ is the circular velocity of the halo, 
$\lambda$ is the spin, $f_c$ is given by equation (\ref{E_binding})
and $f_R$ is a factor taking into account disk action.
If $c$ were a constant, $\Sigma_0$ would decrease 
rapidly with time. This is not natural, because it requires the disk
to re-adjust its structure in the inner region during the 
formation. If, on the other hand, halo concentration increases
with time, as is found for haloes in the slow
accretion phase, the change in the central structure of the 
disk is reduced.  
  
 Our results have also important implications for 
the understanding of the observed Tully-Fisher (TF) relation.
Recent observations show that galaxy disks at modest 
redshifts ($z\sim 1$) obey a TF relation
similar to that for the disk galaxies in the local universe
(e.g. Vogt et al. 1997; Ziegler et al. 2002).
This would be difficult to understand in current theory,
if the halo concentration parameter remained constant, 
because galactic haloes with a given circular velocity
have masses proportional to $H^{-1}(z)$, and so disks
formed at higher redshifts are expected to be lighter
(i.e. have lower TF zero point).
However, as shown in Mo \& Mao (2000), if galaxy 
haloes at high redshifts have lower concentrations, the boost 
of the maximum rotation velocity of the disk relative to the 
halo circular velocity is smaller. This can compensate for
the decrease in the disk mass, and ensure that disks formed at
different redshifts obey roughly the same TF 
relation [see Mo \& Mao (2000) for a more detailed discussion].
    
\section*{acknowledgments}
  We thank the GIF group and the VIRGO consortium for the 
public release of their $N$-body simulation data. We thank 
Frank van den Bosch and Saleem Zaroubi for stimulating discussions.
DHZ thanks the CAS-MPG exchange program for support. 
The research work was supported in part by the
One-Hundred-Talent Program, by NKBRSF (G19990754) and by NSFC
(No.10125314). 




\clearpage

\end{document}